\documentclass[fleqn]{article}
\usepackage{espcrc2}
\usepackage{graphicx}
\usepackage[figuresright]{rotating}
\usepackage{psfig}
\newcommand{\kskl}{K_S^0 K_L^0}
\newcommand{\ks}{K_S^0}

\newcommand{\BR}{{\cal B}}
\newcommand{\eff}{\varepsilon}
\newcommand{\psp}{\psi(2S)}
\newcommand{\jpsi}{J/\psi}
\newcommand{\chicJ}{\chi_{cJ}}
\newcommand{\chicz}{\chi_{c0}}

\newcommand{\chict}{\chi_{c2}}

\newcommand{\EE}{e^+e^-}
\newcommand{\MM}{\mu^+\mu^-}

\newcommand{\pp}{\pi^+\pi^-}

\newcommand{\ppb}{p\overline{p}}
\newcommand{\ksks}{K^0_S K^0_S}

\newcommand{\kskn}{\overline{K}^*(892)^0 K^0 + c.c.}

\newcommand{\jpsipp}{\pi^+\pi^-J/\psi}

\newcommand{\ra}{\rightarrow}
\newcommand{\jpsito}{J/\psi \rightarrow }

\newcommand{\pspto}{\psi(2S) \rightarrow }
\newcommand{\chicJto}{\chi_{cJ} \rightarrow }

\newcommand{\chictto}{\chi_{c2} \rightarrow }
\newcommand{\bfg}{\begin{figure}}
\newcommand{\efg}{\end{figure}}
\newcommand{\bitm}{\begin{itemize}}
\newcommand{\eitm}{\end{itemize}}
\newcommand{\bnum}{\begin{enumerate}}
\newcommand{\enum}{\end{enumerate}}
\newcommand{\btbl}{\begin{table}}
\newcommand{\etbl}{\end{table}}
\newcommand{\btbu}{\begin{tabular}}
\newcommand{\etbu}{\end{tabular}}
\begin{document}

\title{Search for $\ksks$ in $\jpsi$ and $\psp$ decays}
\author{
J.~Z.~Bai$^1$,        Y.~Ban$^{10}$,         J.~G.~Bian$^1$,
X.~Cai$^{1}$,         J.~F.~Chang$^1$,       H.~F.~Chen$^{16}$,
H.~S.~Chen$^1$,       H.~X.~Chen$^{1}$,      J.~Chen$^{1}$,
J.~C.~Chen$^1$
% ...
\\(BES Collaboration)\\
$^1$ Institute of High Energy Physics, Beijing 100039, People's Republic of
     China\\
$^2$ China Center of Advanced Science and Technology, Beijing 100080,
     People's Republic of China\\
$^3$ Guangxi Normal University, Guilin 541004, People's Republic of China\\
$^4$ Henan Normal University, Xinxiang 453002, People's Republic of China\\
$^5$ Huazhong Normal University, Wuhan 430079, People's Republic of China\\
$^6$ Hunan University, Changsha 410082, People's Republic of China\\
$^7$ Liaoning University, Shenyang 110036, People's Republic of China\\
$^{8}$ Nanjing Normal University, Nanjing 210097, People's Republic of China\\
$^{9}$ Nankai University, Tianjin 300071, People's Republic of China\\
$^{10}$ Peking University, Beijing 100871, People's Republic of China\\
$^{11}$ Shandong University, Jinan 250100, People's Republic of China\\
$^{12}$ Shanghai Jiaotong University, Shanghai 200030,
        People's Republic of China\\
$^{13}$ Sichuan University, Chengdu 610064,
        People's Republic of China\\
$^{14}$ Tsinghua University, Beijing 100084,
        People's Republic of China\\
$^{15}$ University of Hawaii, Honolulu, Hawaii 96822\\
$^{16}$ University of Science and Technology of China, Hefei 230026,
        People's Republic of China\\
$^{17}$ Wuhan University, Wuhan 430072, People's Republic of China\\
$^{18}$ Zhejiang University, Hangzhou 310028, People's Republic of China
}

\date{\today}

\begin{abstract}

The CP violating processes $\jpsito \ksks$ and $\pspto \ksks$ are
searched for using samples of 58 million $\jpsi$ and 14 million $\psp$
events collected with the Beijing Spectrometer at the Beijing Electron
Positron Collider.  No signal is observed, and upper limits on the
decay branching ratios are determined to be \( \BR(\jpsito \ksks) <
1.0\times 10^{-6}\) and \( \BR(\pspto \ksks) < 4.6\times 10^{-6}\) at
the 95\% confidence level.

\end{abstract}

\maketitle

\section{Introduction}

The decay of $J^{PC}=1^{--}$ charmonium states,
like $\jpsi$ and $\psp$, to $\ksks$ is a CP violating process. 
However, since CP violations in both the $K^0\overline{K^0}$ and
$B^0\overline{B^0}$ systems have been well established~\cite{cpv},
it is of interest to search for them in other possible 
processes.

Furthermore, it has been shown~\cite{EPR} that the $\ksks$ system can
be used to test the Einstein-Podolsky-Rosen (EPR) paradox versus
quantum theory. The space-like, separated coherent quantum
system may yield some $\jpsito \ksks$ decays if EPR's
locality is correct while quantum theory forbids this
process.

MARK-III searched for decays of $\jpsito \ksks$ with 2.7~million
$\jpsi$ events~\cite{mk3ksks}. No signal was observed, and the
upper limit on the decay branching ratio was
determined to be
$\BR(\jpsito \ksks) < 5.2\times 10^{-6}$
at the 90\% confidence level (C. L.).
The same search for $\psp$ decays has not been performed before.

In this Letter, we report on a search for $\jpsito \ksks$ using a
sample with 20 times more statistics than before, and on the first search for
$\pspto \ksks$. The data samples used for the analyses are taken with
the Beijing Spectrometer (BESII) detector at the Beijing Spectrometer
(BEPC) storage ring at a center-of-mass energies corresponding to
$M_{\jpsi}$ and $M_{\psp}$. The data samples contain $(57.7
 \pm 2.7)\times 10^6$ $\jpsi$ events and $(14 \pm 0.7)\times
10^6$ $\psp$ events, as determined from inclusive hadronic
decays~\cite{fangss,pspscan}.

BES is a conventional solenoidal magnet detector that is
described in detail in Ref.~\cite{bes}; BESII is the upgraded version
of the detector~\cite{bes2}. A 12-layer vertex
chamber (VC) surrounding the beam pipe provides trigger
information. A forty-layer main drift chamber (MDC), located
radially outside the VC, provides trajectory and energy loss
($dE/dx$) information for charged tracks over $85\%$ of the
total solid angle.  The momentum resolution is
$\sigma _p/p = 0.017 \sqrt{1+p^2}$ ($p$ in $\hbox{\rm GeV}/c$),
and the $dE/dx$ resolution for hadron tracks is $\sim 8\%$.
An array of 48 scintillation counters surrounding the MDC  measures
the time-of-flight (TOF) of charged tracks with a resolution of
$\sim 200$ ps for hadrons.  Radially outside the TOF system is a 12
radiation length, lead-gas barrel shower counter (BSC).  This
measures the energies
of electrons and photons over $\sim 80\%$ of the total solid
angle with an energy resolution of $\sigma_E/E=22\%/\sqrt{E}$ ($E$
in GeV).  Outside of the solenoidal coil, which
provides a 0.4~Tesla magnetic field over the tracking volume,
is an iron flux return that is instrumented with
three double layers of  counters that
identify muons of momentum greater than 0.5~GeV/$c$.

\section{Monte Carlo}

A Monte Carlo (MC) simulation is used for the determination of mass
resolutions and detection efficiencies. For the signal channels,
$\jpsito \ksks$ and $\pspto \ksks$, the angular distribution of one
$\ks$ is generated with the same distribution as for the CP allowed
channels $\jpsito \kskl$ and $\pspto \kskl$, namely as $\sin^2\theta$,
where $\theta$ is the polar angle of the $\ks$ in the laboratory
system. Only $\ks \ra \pp$ is generated, and the $\ks$ is allowed to
decay in the detector according to its lifetime. The simulation of the
detector response is done using a Geant3 based program with detailed
consideration of the detector performance (such as dead
electronic channels). Reasonable agreement between data and Monte
Carlo simulation has been observed in various channels tested,
including $\EE \ra (\gamma)\EE$, $\EE\ra (\gamma)\MM$, $\jpsito \ppb$
and $\pspto \jpsipp, \jpsito \ell^+\ell^-$ $(\ell=e,\mu)$.
For this study, 10,000 events for each signal channel are
produced.

\section{Event selection}

For the decay channels of interest, the candidate events are
required to satisfy the following selection criteria:
\begin{enumerate}
\item   The number of charged tracks is required to be four
        with net charge zero. Each track should satisfy
    $|\cos\theta|<0.8$, where $\theta$ is the polar angle
    in the MDC, and should have a good helix
    fit so that the error matrix of the track fitting is
    available for secondary vertex finding.
\item   The tracks are assumed to be either $\pi^+$ or $\pi^-$.
        The higher momentum positive track and the lower momentum
    negative track are assumed to come from one of the $\ks$ decays,
    and the remaining two tracks are assumed to come from the other
    $\ks$ decay. The intersection of the track-pairs
    near the interaction point
    are determined and are regarded as the $\ks$ vertices.

\item  Each $\ks$ candidate is required to have a decay length in the
transverse plane ($L_{xy}$) greater than 3~mm and an invariant mass
within twice the mass resolution as estimated from Monte Carlo
simulations (the mass resolution in $\jpsi$ decays is 7.1~MeV/$c^2$
and in $\psp$ decays 7.9~MeV/$c^2$).
\end{enumerate}

For $\psp$ decays, backgrounds from $\pspto \jpsipp$ with $\jpsi$
decaying into two charged tracks are removed by further requiring
the mass recoiling against the lower momentum positive and negative 
tracks, $m_{rec}$, to be less than 3.0~GeV/$c^2$. The comparison of
the recoil mass distributions of signal and background events
from Monte Carlo simulation is shown in Figure~\ref{mrec}.
It can be seen that this requirement is very efficient in removing
background, and the loss of efficiency is small ($\sim$ 4\%).

\begin{figure}[htbp]
\centerline{\hbox{ \psfig{file=mrec-plb1.epsi,width=6.0cm}}}
\caption{Recoil mass distributions of Monte Carlo simulated
$\pspto \ksks$ (blank histogram) and $\pspto \jpsipp$, $\jpsito
\ell^+\ell^-$ $(\ell=e,\mu)$ (shaded histogram). The histograms
are not normalized. The requirement $m_{rec}<3.0$~GeV/$c^2$ is applied
in the selection of $\psp$ events.} \label{mrec}
\end{figure}

\begin{figure}[htbp]
\centerline{\hbox{ \psfig{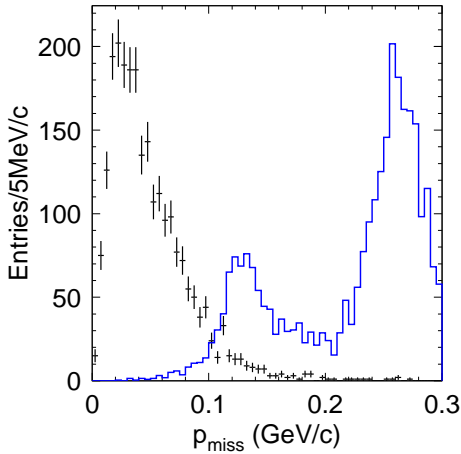}}}
\caption{Missing momentum distributions of Monte Carlo simulated
$\pspto \ksks$ (error bars) and $\pspto \gamma \chicJ$, $\chicJto
\ksks$ ($J=0,2$) (histogram). The histograms are not normalized. An
additional requirement
of $p_{miss}<0.1$~GeV/$c$ is applied to remove the $ \chicJ$
background.} \label{pmiss}
\end{figure}

Another possible background is from $\pspto \gamma \chicJ$ $(J=0,2)$
with $\chicJto \ksks$. Figure~\ref{pmiss} shows $\ksks$ missing
momentum distributions for the signal and  background
channels, all simulated by Monte Carlo. We make an additional
requirement $p_{miss}<0.1$~GeV/$c$ for $\psp$ selection. This removes
94\% of $\chict$ background and almost all the $\chicz$ background,
while the efficiency for $\pspto \ksks$ is about 92\% according to
Monte Carlo simulation.  For $\jpsi$ decays, since $\eta_c \ra \ksks$
is forbidden by parity conservation, no requirement is applied.

The momenta of the two $\ks$ tracks for events passing the above
selection are plotted in Figures~\ref{pkspksj} and \ref{pkspks}, for
$\jpsi$ and $\psp$ decays, respectively, together with the regions
predicted for signal events by the Monte Carlo simulation.  
The two circles correspond to
  \[ \sqrt{(p_{\ks}^a-p_0)^2+(p_{\ks}^b-p_0)^2}<n\sigma, \]
where $p_{\ks}^a$ and $p_{\ks}^b$ are the momenta of the two $\ks$
candidates in the event; $\sigma$ is the $\ks$ momentum resolution,
which is 28~MeV/$c$ in $\jpsi$ decays and 34~MeV/$c$ in $\psp$ decays,
as determined from Monte Carlo simulations; and $p_0$ is the expected
$\ks$ momentum in $\jpsi$ (1.466~GeV) or $\psp$ (1.775~GeV) two-body
decays. The inner circles correspond to $n=1$, and the outer circles
are for $n=2$.  It can be seen that there are no candidate events in
either $\jpsi$ or $\psp$ decays within the one $\sigma$ circle, and
there is only one candidate in each case lying between the one and two
$\sigma$ circles. Using the 2$\sigma$ circle for our event selection,
one event is obtained each for $\jpsito \ksks$ and for $\pspto \ksks$.
The corresponding efficiencies, $\eff_{MC}$, are 20.74\% and 19.18\%
for $\jpsi$ and $\psp$ decays, respectively, as estimated from Monte
Carlo samples.

\begin{figure}[htbp]
\centerline{\hbox{ \psfig{file=pkspks-j-plb1.epsi,width=6.cm}}}
\caption{Momentum of one $\ks$ versus the other for the selected
$\jpsi$ decay candidates (open squares). The circles correspond to one
and two $\sigma$ regions of Monte
Carlo $\jpsito \ksks$ events.}
\label{pkspksj}
\end{figure}

\begin{figure}[htbp]
\centerline{\hbox{ \psfig{file=pkspks-plb1.epsi,width=6.cm}}}
\caption{Momentum of one $\ks$ versus the other for the selected
$\psp$ decay candidates (open squares). The circles correspond to one
and two $\sigma$ regions of Monte
Carlo $\pspto \ksks$ events.}
\label{pkspks}
\end{figure}

The backgrounds remaining in both $\jpsi$ and $\psp$ decays after the
$\ks$ requirement are studied with Monte Carlo simulations. The main
background channel in $\jpsi$ decays is from $\jpsito \kskn$ is
estimated to be $1.0\pm 0.4$ events, which is obtained using a Monte
Carlo sample six times as large as the $\jpsi$ sample. The main
backgrounds channels in $\psp$ decays are $\pspto \gamma \chict$,
$\chictto \ksks$ and $\pspto \kskn$. A Monte Carlo simulation
estimates the total background to be $0.67\pm 0.25$ events, with about
80\% due to $\chictto \ksks$. The errors on the estimated backgrounds
are from the uncertainties of the branching ratios used and the
limited statistics of the Monte Carlo samples.

The two candidate events are investigated further, and it is found
that the $\jpsito \ksks$ candidate is most likely a $\jpsito \kskn$,
with $K^{*0} \ra K^-\pi^+$ misidentified as $\ks \ra \pp$ because no
particle identification is used in the event selection, while the
$\pspto \ksks$ candidate looks like a $\pspto \gamma \chict$,
$\chictto \ksks$ event with an isolated photon with BSC energy of
126~MeV.

\section{Trigger efficiency}

Due to the long decay length of the high momentum $\ks$ and the
requirement of hits in the VC, the trigger efficiency of $\ksks$
events is lower than for normal hadronic events. Since the trigger
system is not included in the Monte Carlo simulation, the trigger
efficiency is measured with data by comparing events within and beyond
the outer radius of the VC using $\jpsi (\psp) \ra \kskl$ events,
which yields trigger efficiencies of $(86.7\pm 0.9)\%$ and $(81.3\pm
1.9)\%$ for a single $\ks$ in $\jpsi$ and $\psp$ decays,
respectively. For $\ksks$ events, the trigger efficiency,
$\eff_{trig}$, is calculated to be $(98.2\pm 0.2)$\% in $\jpsi$ decays
and $(96.5\pm 0.7)$\% in $\psp$ decays.

\section{Secondary Vertex Finding}

The secondary vertex finding algorithm is checked using
$\jpsito K^{*}(892)\overline{K} + c.c.$ events. It is found that the
Monte Carlo simulates data fairly well, and extrapolating the difference
between data and Monte Carlo simulation to the $\ks$ momentum
range under study and correcting by the polar angle dependence of the
efficiency, correction factors of $(96.4\pm 3.1)\%$  and
$(98.1\pm 4.0)\%$ are obtained for a single $\ks$ in $\jpsi$ and $\psp$
decays, respectively. For $\ksks$ decays, the correction factors and
errors to the
Monte Carlo efficiencies, $ \eff_{2nd}$, for $\jpsito \ksks$ and $\pspto
\ksks$ are $(92.9\pm 4.5)\%$ and $(96.2\pm 5.8)\%$, respectively.

\section{Systematic error}

The total systematic error on the branching ratio measurement comes from
all sources listed in Table.~\ref{sys}.
The simulation of the tracking efficiency agrees with data within
1-2\% for each charged track as measured using channels like
$\jpsito \Lambda \overline{\Lambda}$ and $\pspto \pp \jpsi$,
$\jpsito \MM$, and 8\% is taken conservatively as the systematic error
for the channel of interest.

\begin{table}[htbp]
\caption{Summary of the systematic errors.}
\begin{center}
\begin{tabular}{l|cc}
\hline\hline
Source                      & $\jpsi$ (\%) & $\psp$(\%) \\\hline
MC statistics               &  2.1       &  2.0   \\
Trigger efficiency          &  0.2       &  0.7   \\
$2^{nd}$ vertex             &  4.8       &  6.0   \\
MDC tracking                &  8.0       &  8.0   \\
Resolutions                 &  5.4       &  5.6   \\
Number of events            &  4.7       &  5.0    \\
$\BR(\ks \ra \pp)$          &  0.8       &  0.8     \\
\hline
Sum                         & 12.0       & 12.7   \\\hline\hline
\end{tabular}
\end{center}
\label{sys}
\end{table}

The Monte Carlo simulated $\ks$ mass and momentum resolutions
agree with those of data within statistical uncertainties,
as has been checked with $\jpsi$ and $\pspto \kskl$
samples~\cite{bes2ksklj,bes2kskl}. The requirement that the $\ks$ mass and
momentum be within two standard deviations introduces systematic
uncertainties at the 5-6\% level, dominated by the statistical
precisions of the comparisons between data and Monte Carlo
simulation.

The systematic errors on the $p_{miss}$ and the recoil mass requirements
for the $\psp$ sample depends on the simulation of the momentum
of the charged tracks and are already included either in the tracking
or in the systematic error quoted for the $\ks$ mass and momentum
resolutions. They are not further considered here.

The systematic errors on the total
numbers of $\jpsi$ and $\psp$ events are taken as
4.7\% and 5.0\%, respectively, and are
measured using inclusive hadronic events with four charged hadrons
in the final state~\cite{fangss} for $\jpsi$ and using inclusive hadrons for
$\psp$~\cite{pspscan}.
The systematic error on the branching ratio used,
$\BR(\ks \ra \pp)$ is obtained from the Particle 
Data Group~\cite{pdg} directly.
Adding all the systematic errors in quadrature, the total
systematic errors are 12.0\% and 12.7\% for $\jpsi$ and
$\psp$ decays, respectively.

\section{Results}

Since the observed numbers of events agree with the expected
background levels for the channels under study, upper limits are
conservatively set by not subtracting background and taking the events in
MC predicted region as signal. With one observed event, the upper
limit on the number of events is 4.74 at the 95\% confidence level.

The upper limits on the branching ratios of $\jpsi$ and
$\pspto \ksks$ are calculated with
\[ \BR(R \to \ksks)<
    \left.\frac{n^{obs}_{UL}/(\eff_{MC}\cdot \eff_{trig}\cdot \eff_{2nd})}
               {N_{R}\cdot \BR(\ks\ra \pp)^2}
     \right. , \]
where $n^{obs}_{UL}$ is the upper limit of the number of observed
events, and $N_{R}$ is the total number of resonance $R$ events.
Using the numbers listed in Table~\ref{br}, one
obtains the upper limits on the branching ratios at the 95\% C. L.:
  \[  \BR(\jpsito \ksks)<1.0\times 10^{-6}, \]
  \[  \BR(\pspto  \ksks)<4.6\times 10^{-6}, \]
where the systematic errors are included by lowering the
efficiencies by one standard deviation.

\begin{table}[htbp]
\caption{Numbers used in the calculations of upper limits and the
final results.}
\begin{center}
\begin{tabular}{l|cc}
\hline\hline
     $R$                  & $\jpsi$           & $\psp$ \\\hline
$n^{obs}$                 & $1$               &   $1$ \\
$n^{obs}_{UL}$            & $4.74$            &   $4.74$ \\
$\eff_{MC}$ (\%)          & $20.74 \pm 0.41$  & $19.18\pm 0.39$\\
$\eff_{trg}$ (\%)         & $98.2 \pm 0.2 $   & $96.5\pm 0.7$\\
$\eff_{2nd}$ (\%)         & $92.9 \pm 4.5 $   & $96.2\pm 5.8$\\
$N_{\psp} (10^6)$         & $57.7\pm 2.7  $   & $14.0\pm 0.7$ \\
$\BR(\ks \ra \pp)$        & \multicolumn{2}{c}{$0.6860\pm 0.0027$} \\\hline
$\BR(R\to \ksks)<$        & $1.0\times 10^{-6}$ & $4.6\times 10^{-6}$   \\
\hline\hline
\end{tabular}
\end{center}
\label{br}
\end{table}

\section{Summary}

The CP violating processes $\jpsito \ksks$ and $\pspto \ksks$ are
searched for using the BESII 58 million $\jpsi$ event and 14 million
$\psp$ event samples. The upper
limits on the branching ratios are determined to be
\( \BR(\jpsito \ksks) <1.0\times 10^{-6} \) and \( \BR(\pspto
\ksks)  <4.6\times 10^{-6} \) at the 95\% C. L. 
The former is much more stringent
than the previous MARK-III measurement~\cite{mk3ksks}, and the
latter is the first search for this channel in $\psp$ decays. Current
bounds of the production rates still beyond the sensitivity for
testing the EPR paradox~\cite{EPR}.

\section*{Acknowledgments}

We wish to thank Dr. H.~B.~Li for helpful comments.
The BES collaboration thanks the staff of BEPC for their
hard efforts. This work is supported in part by the National
Natural Science Foundation of China under contracts
Nos. 19991480, 10225524, 10225525, the Chinese Academy
of Sciences under contract No. KJ 95T-03, the 100 Talents
Program of CAS under Contract Nos. U-11, U-24, U-25, and
the Knowledge Innovation Project of CAS under Contract
Nos. U-602, U-34(IHEP); by the National Natural Science
Foundation of China under Contract No. 10175060 (USTC);
and by the Department of Energy under Contract
No. DE-FG03-94ER40833 (U Hawaii).


\begin{thebibliography}{**}

\bibitem{cpv} A summary on the CP violation in Kaon and B decays
              can be found in
          K.~Hagiwara {\em et al.} (Particle Data Group),
              Phys. Rev. {\bf D66}, 010001-118 (2002), and
          references therein.
\bibitem{EPR} M.~Roos, ``Test of Einstein Locality'',
              HU-TFT-80-5 (revised), Nov. 1980,
          and references there in.

\bibitem{mk3ksks} R.~M.~Baltrusaitis {\em et al.},
                Phys. Rev. {\bf D32}, 566 (1985).

\bibitem{fangss} S.~S.~Fang {\em et al.},
         HEP\&NP {\bf 27}, 277 (2003) (in Chinese).

\bibitem{pspscan} J.~Z.~Bai. {\em et al.} (BES Collab.),
             Phys. Lett. {\bf B550}, 24 (2002).

\bibitem{bes} J.~Z.~Bai. {\em et al.} (BES Collab.), Nucl. Instr. Meth.
              {\bf A344}, 319 (1994).

\bibitem{bes2} J.~Z.~Bai. {\em et al.} (BES Collab.), Nucl. Instr. Meth.
              {\bf A458}, 627 (2001).

\bibitem{bes2ksklj} J.~Z.~Bai. {\em et al.} (BES Collab.),
              hep-ex/0310023, to appear in Phys. Rev. {\bf D}.

\bibitem{bes2kskl} J.~Z.~Bai. {\em et al.} (BES Collab.),
              hep-ex/0310024, to appear in Phys. Rev. Lett.

\bibitem{pdg} K.~Hagiwara {\em et al.} (Particle Data Group),
              Phys. Rev. {\bf D66}, 010001 (2002).

\end{thebibliography}
\end{document}